\documentclass[lineno]{jfm}

\usepackage{graphicx}
\usepackage{newtxtext}
\usepackage{newtxmath}
\usepackage{natbib}
\usepackage{hyperref}
\usepackage{bm}
\hypersetup{
    colorlinks = true,
    urlcolor   = blue,
    citecolor  = black,
}

\newcommand{\RomanNumeralCaps}[1]
\linenumbers

\title{Scale-resolved turbulent Prandtl number for Rayleigh-B\'{e}nard convection at $\Pran\boldsymbol{=10^{-3}}$}

\author{Shashwat Bhattacharya\aff{1}
  \corresp{\email{shashwat@iitmandi.ac.in}},
        Dmitry Krasnov\aff{2},
        Ambrish Pandey\aff{3},\\ 
        Toshiyuki Gotoh\aff{4,5}\and 
        J{\"o}rg Schumacher \aff{2}}

\affiliation{
\aff{1}School of Mechanical and Materials Engineering, Indian Institute of Technology Mandi, Kamand 175005, India
\aff{2}Institute of Thermodynamics and Fluid Mechanics, Technische Universit{\"a}t Ilmenau, P.O.Box 100565, D-98684 Ilmenau, Germany
\aff{3}Department of Physics, Indian Institute of Technology Roorkee, Roorkee 247667, India
\aff{4}Department of Engineering, Nagoya Institute of Technology, Nagoya 466-8555, Japan
\aff{5}Research and Education Center for Natural Sciences, Keio University, Hiyoshi, Yokohama, 223-8521, Japan}

\begin{document}
\maketitle

\begin{abstract}
We present a framework to calculate the scale-resolved turbulent Prandtl number $\Pran_t$ for the well-mixed and highly inertial bulk of a turbulent Rayleigh-B\'{e}nard mesoscale convection layer at a molecular Prandtl number $\Pran=10^{-3}$. It builds on Kolmogorov's refined similarity hypothesis of homogeneous isotropic fluid and passive scalar turbulence, based on log-normally distributed amplitudes of kinetic energy and scalar dissipation rates that are coarse-grained over variable scales $r$ in the inertial subrange. Our definition of turbulent (or eddy) viscosity and diffusivity does not rely on mean gradient-based Boussinesq closures of Reynolds stresses and convective heat fluxes. Such gradients are practically absent or indefinite in the bulk. The present study is based on direct numerical simulation of plane-layer convection at an aspect ratio of $\Gamma=25$ for Rayleigh numbers $10^5\le Ra\le 10^7$. We find that the turbulent Prandtl number is effectively up to 4 orders of magnitude larger than the molecular one, $\Pran_t\sim 10$. This holds particularly for the upper end of the inertial subrange, where the eddy diffusivity exceeds the molecular value, $\kappa_e(r)>\kappa$. Highly inertial low-Prandtl-number convection behaves effectively as a high-Prandtl number flow, which also supports previous models for the prominent application case of solar convection.
\end{abstract}

\section{Introduction}
\label{sec:Introduction}

The turbulent Prandtl number $\Pran_t$ is an essential parameter in many turbulence applications, particularly for astrophysical or geophysical examples with extreme amplitudes of their dimensionless control parameters, such as in solar and stellar convection~\citep{Schumacher:RMP2020,Kapyla:AA2021,Garaud:PRF2021,Alam2025}. It is required to construct accurate turbulence models for simulating highly turbulent heat transport in the Sun and other stars, or in the atmosphere \citep{Li:AR2019}, along with various engineering applications, such as cooling liquid metal blankets in nuclear reactors~\citep{Otic:NSE2007}.  The turbulent Prandtl number is given as the ratio of the turbulent viscosity $\nu_e$ to the turbulent diffusivity $\kappa_e$:
\begin{equation}
\Pran_t=\frac{\nu_e}{\kappa_e}\,.
\end{equation}
The values of $\nu_e$ and $\kappa_e$ can differ by orders of magnitude from their molecular counterparts, the molecular viscosity $\nu$ and the molecular diffusivity $\kappa$. Thus, they shift the range of $\Pran_t$ significantly away from that of the molecular value $\Pran = \nu/\kappa$. For moderate Prandtl number fluids at $\Pran \sim 1$, such as air and water, $\Pran_t$ is predicted with reasonable accuracy using the Reynolds analogy, according to which the eddies that are responsible for the turbulent transport of momentum are also responsible for transporting heat~\citep{Bricteux:NED2012,Li:AR2019,Abe:IJHMT2019}. This argument implies that $\Pran_t \approx 1$. \citet{Rudiger1989} derived turbulent Prandtl numbers of $\Pran_t\sim {\cal O}(1)$ by analytical mean field calculations in Fourier spectral space for $0.01\lesssim \Pran\lesssim 1$. It must be noted that Reynolds' analogy does not hold true for liquid metals ($\Pran \ll 1$), for which $\Pran_t$ has been observed to be larger than unity~\citep{Reynolds:IJHMT1975,Jischa:IJHMT1979,Bricteux:NED2012}. Similar observations have been made for passive scalars by \cite{Donzis:JFluidsEngineering2005} for the turbulent Schmidt number in case of passive scalar mixing. Recent investigations on turbulent channel flow~\citep{Abe:IJHMT2019} and thermal convection~\citep{Tai:PRF2021,Pandey:PRF2021,Pandey:PD2022} suggest that $\Pran_t$ increases with decreasing $\Pran$. Turbulent Prandtl numbers that deviate strongly from unity can have a significant impact on the range of scales which have to be resolved in simulation models \citep[see, for example][]{Bekki:ApJ2017,Karak:PoF2018,Pandey:PRF2021,Pandey:PD2022}. In both parameters, $\nu_e$ and $\kappa_e$, the effect of the turbulent fluctuations at the smallest unresolvable scales is condensed. These fluctuations are typically highly intermittent; however, they are not incorporated in derivations of $\Pran_t$. 

Turbulent viscosity and diffusivity can be obtained by the so-called Boussinesq or flux-gradient ansatz for the turbulent stresses, which is given by 
\begin{equation}
\langle u_i'u_j'\rangle \approx -\nu_e \left(\frac{\partial \langle u_i\rangle}{\partial x_j}+\frac{\partial \langle u_j\rangle}{\partial x_i}\right)\quad \mbox{and} \quad \langle u_i'T'\rangle \approx -\kappa_e \frac{\partial \langle T\rangle}{\partial x_i}\,,
\label{eq:bouss}
\end{equation}
with $i,j \in \lbrace x,y,z \rbrace$. In equations~\eqref{eq:bouss}, the Reynolds decompositions $u_i = \langle u_i \rangle + u_i'$ and $T= \langle T \rangle + T'$ are used, where $\langle \cdot \rangle$ represents the mean and the prime represents the corresponding fluctuation about the mean. These relations rely on the existence of monotonic mean gradient profiles in the flow at hand. \cite{Kapyla:JFM2022} used the flux-gradient ansatz to analyse turbulent Prandtl numbers for $0.01\le \Pran\le 10$, and obtained $\Pran_t\sim 1$ for bulk turbulence, which is sustained by forcing a sinusoidal shear mode that does not affect the local isotropy, but generates a finite mean gradient. Such gradients are absent in Rayleigh-B\'{e}nard convection (RBC) for the mean velocity components, as shown recently in \cite{Samuel:JFM2024}. Other examples are transiently growing convective boundary layers \citep{Heyder:JAMES2024}, which have a zero mean temperature gradient and require additional mass-flux parametrizations for the buoyancy flux \citep{Siebesma:JAS2007}. 

The RBC setup, which is of interest for the present discussion, consists of a fluid enclosed between two parallel horizontal plates of vertical distance $H$, with the bottom plate kept at a higher temperature than the top plate. Recently, \citet{Pandey:PRF2021,Pandey:PD2022} computed $\Pran_t$ for a wide range of Prandtl numbers in RBC using the $k_u$-$\epsilon$ framework, where $k_u$ is the turbulent kinetic energy and $\epsilon$ is the viscous dissipation rate.  However, the study assumed that the ratio of the unknown prefactors of $\nu_e$ and $\kappa_e$ is constant. The validity of this assumption has not yet been verified.

In this paper, we examine the turbulent Prandtl number of RBC at a very low molecular Prandtl number of $\Pran=10^{-3}$. We calculate $\Pran_t$ on the basis of the refined similarity theory, which was developed by \citet{Kolmogorov:JFM1962} and ~\citet{Obukhov:JFM1962} for homogeneous isotropic turbulence and is known under the abbreviation K62.  We consider the bulk of an extended turbulent mesoscale convection layer, where the active character of the temperature field as a driver of fluid motion becomes less important in comparison to that in the boundary layers close to the bottom and top walls. Therefore, we can apply an extension of the K62 framework, which was developed by \citet{Stolovitzky:JFM1995} for passively advected scalar fields in a turbulent flow.  Our analysis is based on the data from high-resolution direct numerical simulations (DNS), obtained by solving the equations for RBC in a horizontally extended layer. Our approach, therefore, minimizes the usage of assumptions and, for the first time, makes the determination of $\Pran_t$ at different scales possible.

The outline of the manuscript is as follows. In \S~2 we briefly describe the derivation of the expressions for the eddy viscosity and diffusivity. \S~3 compactly summarizes the numerical simulations and the data records. This is followed by \S~4 with the results. The manuscript closes with a summary and an outlook. 

\section{Eddy viscosity and diffusivity from refined similarity framework}
\label{sec:RefinedSimilarity}

 In the following, we will derive the expressions for the eddy viscosity, eddy diffusivity, and turbulent Prandtl number using the refined similarity hypothesis.
\subsection{Eddy viscosity from velocity field}
\label{sec:RefinedSimilarityVelocity}
At the core of the theory is the kinetic energy dissipation rate field, which is given by
\begin{equation}
    \epsilon(\boldsymbol{x},t)=\frac{\nu}{2}\sum_{i,j} \left ( \frac{\partial u_i}{\partial x_j} + \frac{\partial u_j}{\partial x_i} \right )^2,
    \label{eq:LocalViscDiss}
\end{equation}
where $u_i$ are the components of the turbulent velocity field. Note that in RBC, ${\bm u}={\bm u}'$.  Based on equation~\eqref{eq:LocalViscDiss}, one can define the following coarse-grained dissipation field:
\begin{equation}
    \epsilon_r(\boldsymbol{x},t)=\frac{1}{B_r} \int_{B_r} \epsilon(\boldsymbol{x}+\boldsymbol{y},t) \, \mathrm{d}^3 y,
    \label{eq:CGViscDiss}
\end{equation}
 where $B_r$ is a sphere of radius $r$ centered at $\boldsymbol{x}$. The coarse-graining scale $r$ varies between the viscous and outer scales of the inertial range of turbulence, which will be Kolmogorov scale $\eta$ and $H$. The logarithm of the new energy dissipation field $\epsilon_r(\boldsymbol{x},t)$ is assumed to be normally distributed. Let us associate a characteristic velocity ${\cal U}_r=(r \epsilon_r)^{1/3}$ and a characteristic time ${\cal T}_r=(r^2/\epsilon_r)^{1/3}$ with the scale $r$, which is taken in the inertial subrange~\citep{Sreenivasan:ARFM1997}. Furthermore, we define the following dimensionless relative velocities~\citep{Kolmogorov:JFM1962}
\begin{equation}
    {\bm w}(\boldsymbol{\xi},t)= \frac{\boldsymbol{u}(\boldsymbol{x}+\boldsymbol{\xi}r, \, t+\tau {\cal T}_r) - \boldsymbol{u}(\boldsymbol{x},t)}{{\cal U}_r}.
    \label{eq:DimenRelVel}
\end{equation}
In the case of assumed statistical stationarity, the temporal arguments can be neglected for the subsequent discussion. The focus is on spatial scales which are locally isotropic. Thus, we set $\boldsymbol{\xi}=(1,0,0)$. A longitudinal velocity increment along this direction follows then with the definition (\ref{eq:LocalViscDiss}) to
\begin{equation}
    \delta u (r) = u_x(\boldsymbol{x} + \boldsymbol{\xi} r, t) - u_x (\boldsymbol{x},t) = w_x(1,0,0,0)(r \epsilon_r)^{1/3}.
    \label{eq:VelIncrement}
\end{equation}
A scale-resolved eddy viscosity in the inertial subrange is then given by
\begin{equation}
    \nu_e(r) \approx r \, \delta u(r) = w_x(1,0,0,0)\,r^{4/3}\epsilon_r^{1/3}.
    \label{eq:ScaleEViscosity}
\end{equation}
In the following, we simply write $w_x$ and drop the argument vector. The distribution of the fluctuating prefactor is unknown and has to be determined from DNS. This was done in \citet{Iyer:PRE2017} for high-Reynolds number data for isotropic box turbulence. The probability density function (PDF) of $w_x$ was found to obey super-Gaussian tails. Furthermore, $\langle w_x \rangle = 0$ and $\langle w_x^3 \rangle = -4/5$ in correspondence with the $4/5$ths law~\citep{Kolmogorov:DANS1941Structure}. The ansatz (\ref{eq:ScaleEViscosity}) for $\nu_e(r)$ is consistent because
\begin{equation}
    \nu_e(\eta)=\lim_{r \rightarrow \eta} \nu_e(r)=\nu \quad \mbox{since} \quad \nu_e(\eta)=\eta \, \delta u (\eta) = \Rey_\eta \nu,
    \label{eq:LimNue}
\end{equation}
with Kolmogorov scale $\eta=(\nu^3/\langle \epsilon \rangle)^{1/4}$ and $\Rey_\eta=1$~\citep{Yakhot:PD2006}. The quantity $\nu_e(r)$ is a highly fluctuating field due to the fluctuations of $\eta$ and $w_x$. It is also dimensionally consistent with the $k_u$-$\epsilon$ type ansatz for $\nu_e$~\citep{Pope:book} which is given by $\nu_e \sim k_u^2/\langle \epsilon \rangle$ with the turbulent kinetic energy $k_u=\langle u^2 \rangle /2$ and $\langle\epsilon\rangle$ being the mean kinetic energy dissipation rate. An \textit{effective eddy viscosity at scale} $r$ in the inertial range follows in the present framework by an ensemble average
\begin{equation}
    \langle \nu_e(r) \rangle = \langle w_x \epsilon_r^{1/3} \rangle r^{4/3}.
    \label{eq:EffectiveScaleEViscosity}
\end{equation}
This moment is to be determined from the simulation data for $\eta \ll r \ll H$, with $H$ being the (scale) height of the turbulent convection layer.

\subsection{Eddy diffusivity from temperature field}
\label{sec:ScalarFields}

In the next step, we extend this framework to the temperature field $T$, inspired by the work of ~\citet{Stolovitzky:JFM1995}. The procedure is a bit less straightforward compared to that for the eddy viscosity. The scalar field at hand is the fluctuating temperature field which is given by
\begin{equation}
    T'(\boldsymbol{x},t)=T(\boldsymbol{x},t)-\langle T(z) \rangle,
    \label{eq:FlucTempField}
\end{equation}
where $\langle T(z) \rangle$ is the area- and time-averaged temperature. The corresponding thermal dissipation rate is given by
\begin{equation}
    \chi(\boldsymbol{x},t) = \kappa \left ( \frac{\partial T'}{\partial x_j} \right )^2.
    \label{eq:ThermalDiss}
\end{equation}
Similar to (\ref{eq:CGViscDiss}), we define the coarse-grained thermal dissipation field:
\begin{equation}
    \chi_r(\boldsymbol{x},t)=\frac{1}{B_r} \int_{B_r} \chi(\boldsymbol{x}+\boldsymbol{y},t) \, \mathrm{d}^3y.
    \label{eq:CGThermalDiss}
\end{equation}
The coarse-graining scale $r$ varies again between the diffusive and outer scales of the inertial-convective range of scalar turbulence, which will be the Corrsin scale $\eta_C$ and $H$. See section 3 for definition.  Following \citet{Stolovitzky:JFM1995}, the increment of temperature fluctuations is given by
\begin{equation}
    \delta T' (r) = T'(\boldsymbol{x} + \boldsymbol{\xi}r,t) - T'(\boldsymbol{x},t) = w_T(1,0,0,0) \frac{(r \chi_r)^{1/2}}{(r \epsilon_r)^{1/6}}.
    \label{eq:TempInc}
\end{equation}
Thus, we get the following expression for the eddy diffusivity, cf. \eqref{eq:ScaleEViscosity}
\begin{equation}
    \kappa_e(r) \approx \frac{[\delta T'(r) \, \delta u (r)]^2}{\chi_r}=w_x^2(1,0,0,0) w_T^2 (1,0,0,0)\, r^{4/3} \epsilon_r^{1/3}.
    \label{eq:ScaleEDiff}
\end{equation}
We note again that the middle term of this equation is consistent with the $k_u$-$\epsilon$ type ansatz for an eddy diffusivity, $\kappa_e \sim k_u k_T/\langle \chi \rangle$ with the thermal variance $k_T=\langle T'^2 \rangle/2$.  The \textit{effective eddy diffusivity at scale} $r$ follows
\begin{equation}
    \langle \kappa_e(r) \rangle = \langle w_x^2 w_T^2 \epsilon_r^{1/3} \rangle r^{4/3}.
    \label{eq:EffectiveEddyVisc}
\end{equation}
We end up with the following expression for the scale-dependent turbulent Prandtl number
\begin{equation}
    \Pran_t(r) = \frac{\langle \nu_e(r) \rangle}{\langle \kappa_e (r) \rangle} = \frac{\langle w_x \epsilon_r^{1/3} \rangle}{\langle w_x^2 w_T^2 \epsilon_r^{1/3}\rangle}.
    \label{eq:ScaleTurbPr}
\end{equation}

We note that the thermal dissipation rate disappears in this derivation for the eddy diffusivity and turbulent Prandtl number. Further, the statistics of the dimensionless $w_T$ will depend on the molecular Prandtl number. So, it can be expected that the $\Pran$-dependence of $\Pran_t$ enters dominantly via this property. In this paper, we investigate how these definitions work for a very low Prandtl number fluid.

\section{Numerical simulation data} \label{sec:Numerical_Method}
To examine the framework developed in \S~\ref{sec:RefinedSimilarity}, we use data obtained from DNS of RBC conducted by \citet{Pandey:JFM2022}. The convection flow domain consists of a closed rectangular box with an aspect ratio of $25H:25H:H$.
The molecular Prandtl number is set at $\Pran=10^{-3}$, and the Rayleigh number $Ra=\{10^5, 10^6, 10^7\}$. The simulations were performed using a second-order finite difference solver developed by \citet{Krasnov:CF2011,Krasnov:JCP2023}. The dimensionless equations of motion are given by
\begin{eqnarray}
\nabla \cdot \boldsymbol{u}&=&0 \label{eq:continuity} \, , \\
\frac{\partial \boldsymbol{u}}{\partial t}  + \boldsymbol{u}\cdot \nabla \boldsymbol{u} &=& -\nabla p + T\hat{z}+ \sqrt{\frac{\Pran}{\Ray}} \nabla^2 \boldsymbol{u}, \label{eq:Momentum} \\
\frac{\partial T}{\partial t} + \boldsymbol{u} \cdot \nabla T &=& \frac{1}{\sqrt{\Ray \, \Pran}} \nabla^2 T, \label{eq:T_energy} 
\end{eqnarray}
where $\boldsymbol{u}$, $p$, and $T$ are the fields of velocity,  pressure, and temperature, respectively. The governing equations are made dimensionless by using the cell height $H$, the imposed temperature difference $\Delta$, and the free-fall velocity $U_f = \sqrt{\alpha g \Delta H}$ (where $g$ and $\alpha$ are, respectively, the acceleration due to gravity and the volumetric coefficient of thermal expansion of the fluid) as the length, temperature, and velocity scales, respectively. The nondimensional governing parameters are the Rayleigh number $\Ray=\alpha g \Delta H^3/(\nu\kappa)$ and the Prandtl number $\Pran=\nu/\kappa$. 

The temperatures of the top and bottom walls were held fixed at $T=-0.5$ and $T=0.5$ respectively, and the sidewalls were adiabatic with $\partial T/\partial n =0$ (where $n$ is the component normal to sidewall).  No-slip boundary conditions were imposed on all the walls. The flow domain was divided into $N_x \times N_y \times N_z$ gridpoints. The mesh was non-uniform in the $z$-direction with stronger clustering of the grid points near the top and bottom boundaries. The elliptic equations for pressure and temperature were solved based on applying cosine transforms in $x$- and $y$-directions and using a tridiagonal solver in the $z$-direction. The diffusive term in the temperature transport equation was treated implicitly. A fully explicit Adams-Bashforth/Backward-Differentiation method of second order~\citep{Peyret:book} was used for time discretization.

The simulations were run for 40 free-fall time units for $\Ray=10^5$ and 10 free-fall time units for $\Ray=10^6$ and $10^7$ after relaxation into a statistically steady state. A full snapshot is saved every two free-fall time units for $\Ray=10^5$ and every free-fall time unit for the remaining cases. A minimum of 13 points was present in the viscous boundary layers, implying that the boundary layers were adequately resolved~\citep{Grotzbach:JCP1983,Verzicco:JFM2003,Shishkina:NJP2010}. 
Here, we define the viscous boundary layer as the region between the wall and the local maximum of the root mean square velocity profile~\citep{Samuel:JFM2024}.
Further resolution tests were provided in the appendix of \citet{Pandey:JFM2022}. 

Important parameters of the simulations are summarized in table~\ref{table:Simulation}.
The global heat transport is quantified using the  Nusselt number, which is computed as $Nu=1 + \sqrt{Ra Pr} \, \langle u_z T\rangle$. Due to very high thermal diffusivity of the fluid, the molecular diffusion contributes significantly in the heat transport, resulting in relatively low values of $Nu$, see table~\ref{table:Simulation}. Next to the Kolmogorov scale $\eta$, we compute the Corrsin scale $\eta_C=\eta/Pr^{3/4}$, which is the finest scale in the temperature field for $\Pran < 1$. Following \cite{Scheel:NJP2013}, we obtain $\eta=\langle\epsilon\rangle^{-1/4} (\Pran/\Ray)^{3/8}$ in dimensionless form. In the present work, the Corrsin length scale is nearly 178 times above the Kolmogorov length scale.
\begin{table}
\captionsetup{width=1\textwidth}
  \begin{center}
\def~{\hphantom{0}}
  \begin{tabular}{lcccccc}
      $\Ray$  & $N_x \times N_y \times N_z$   &  Time frames & $N_{BL}$ & $\Nus$ & $\eta$ & $\eta_C$\\[3pt]
       $10^5$   & $9600 \times 9600 \times 640$ &20 & 18 & $1.21 \pm 0.005$ & $2.63 \times 10^{-3}$ & $4.67 \times 10^{-1}$\\
       $10^6$   & $12800 \times 12800 \times 800$ &10 & 13 & $2.48 \pm 0.005$ & $9.07 \times 10^{-4}$ & $1.61 \times 10^{-1}$\\
       $10^7$  &$20480 \times 20480 \times 1280$ &10 & 13 & $4.57 \pm 0.01$ & $4.09 \times 10^{-4}$ & $7.27 \times 10^{-2}$
  \end{tabular}
  \caption{Details of the direct numerical simulations of RBC with Prandtl number $\Pran=10^{-3}$. The Rayleigh number ($\Ray$), grid size ($N_x \times N_y \times N_z$), the number of saved time frames, the number of grid points in the viscous boundary layer ($N_{BL}$), the Nusselt number ($\Nus$), the Kolmogorov length scale ($\eta$), and the Corrsin length scale ($\eta_C$) are provided.}
  \label{table:Simulation}
  \end{center}
\end{table}

\section{Dissipation rate field statistics}

\begin{figure}
\captionsetup{width=1\textwidth}
  \centerline{\includegraphics[scale=0.5]{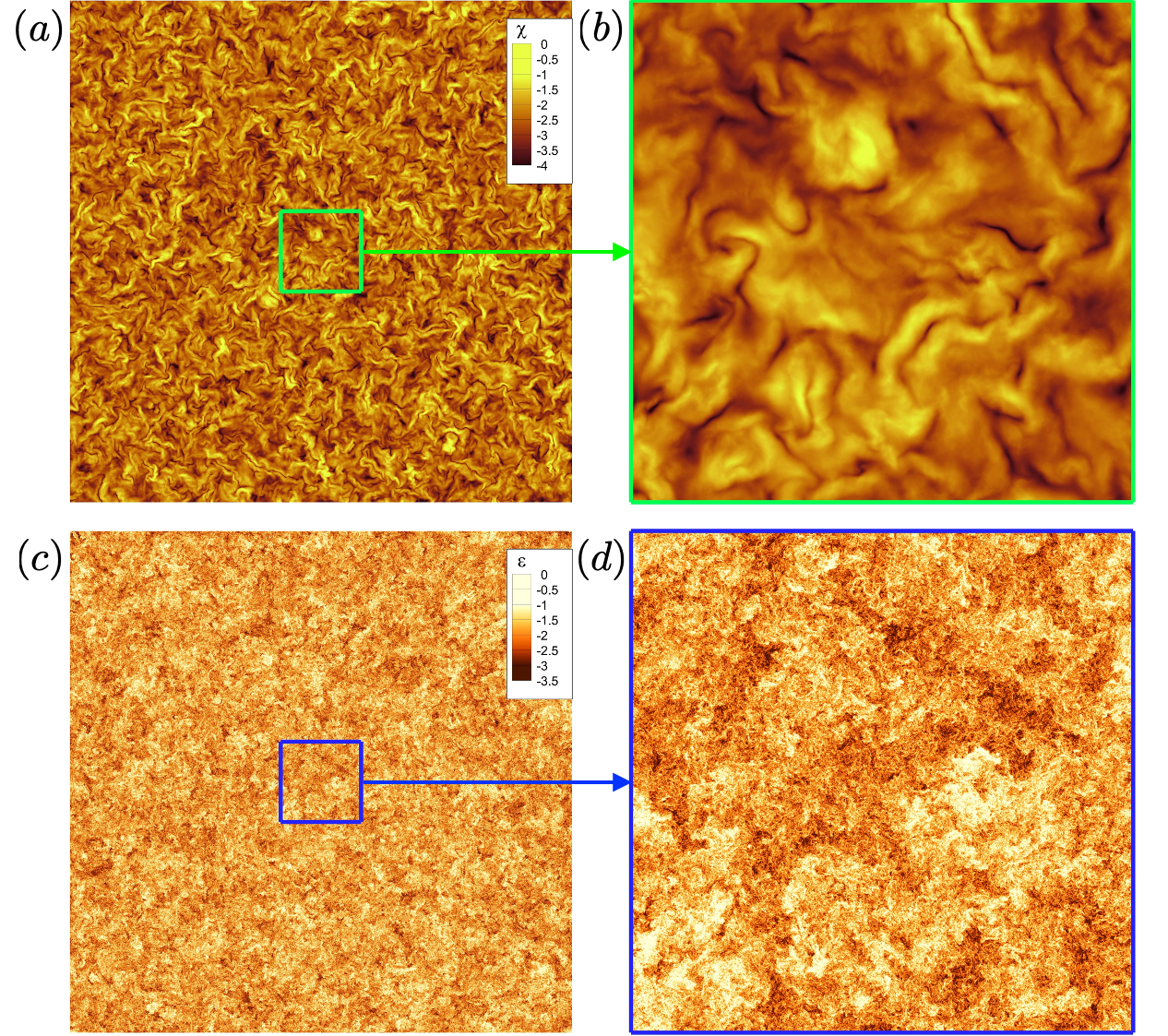}}
  \caption{Contour plots of the dissipation rates in the horizontal midplane for $\Ray=10^7$. (a) Contours of the logarithm of the thermal dissipation rate $\chi$ in the full cross section plane. (b) Magnified image of the region enclosed by the green square in panel (a). (c) Contours of the logarithm of the viscous dissipation rate $\epsilon$ in the full cross section plane. (d) Magnified image of the region enclosed by the blue square in panel (c). The plots show that the characteristic length scale of $\chi$ is much larger than that of $\epsilon$, consistent with the fact that the Corrsin scale $\eta_C$ is much larger than the Kolmogorov length scale $\eta$.}
\label{fig:Dissipation_contours}
\end{figure}

Figure~\ref{fig:Dissipation_contours} displays contour plots of the instantaneous thermal and viscous dissipation rates in the horizontal midplane of the box for the highest accessible Rayleigh number, $\Ray=10^7$. The viscous dissipation rate field $\epsilon$ exhibits much finer structures than the thermal dissipation rate which is in line with $\eta \ll \eta_C$, as listed in table \ref{table:Simulation}. 

It must be recalled that the derivation of the expressions for the scale-resolved eddy viscosity and diffusivity, and thus the turbulent Prandtl number in \S~\ref{sec:RefinedSimilarity}, follows from the refined self-similarity framework of \citet{Kolmogorov:JFM1962}, which extended the classical turbulence theory and included fluctuations of the disspation fields. K62 assumes that the logarithms of the coarse-grained thermal and viscous dissipation rates follow a normal distribution for $r \ll H$. We will now verify this assumption using our DNS data. To this end, we compute the coarse-grained viscous dissipation rate $\epsilon_r$ and thermal dissipation rate $\chi_r$ using equations~\eqref{eq:LocalViscDiss}--\eqref{eq:CGViscDiss} and \eqref{eq:ThermalDiss}--\eqref{eq:CGThermalDiss}, respectively. The scales, at which the increments are taken, are $r=3\eta$, $9\eta$, and $100\eta$ for the viscous dissipation rate, and $r=\eta_C$, $2 \eta_C$, and $4 \eta_C$ for the thermal dissipation rate. 

\begin{figure}
\captionsetup{width=1\textwidth}
  \centerline{\includegraphics[scale=0.4]{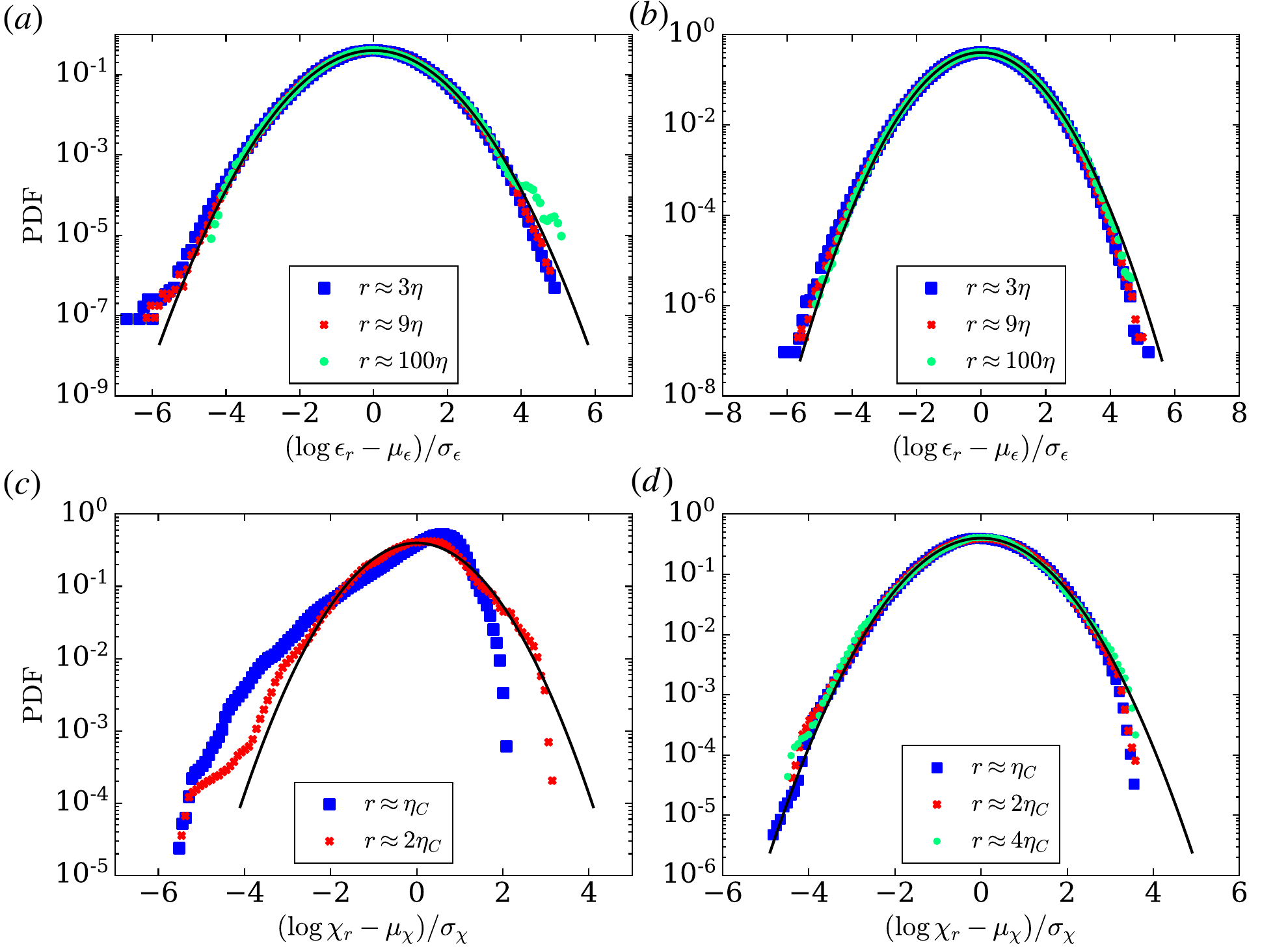}}
  \caption{Probability density functions (PDFs) of normalized viscous and thermal dissipation rates. The top row exhibits the PDFs of $\mathrm{log}~\epsilon_r$, centered with respect to its mean $\mu_\epsilon$ and normalized by its standard deviation $\sigma_\epsilon$, with $3 \eta \leq r \leq 100 \eta$. Panel (a) is for $\Ray=10^5$ and (b) for $\Ray=10^7$. The bottom row exhibits the PDFs of $\mathrm{log}~\chi_r$, normalized with respect to its mean $\mu_\chi$ and its standard deviation $\sigma_\chi$, with $\eta_C \leq r \leq 4 \eta_C$. Panel (c) is for $\Ray=10^5$ and (d) for $\Ray=10^7$. For $\epsilon_r$, the PDFs collapse reasonably well with a Gaussian curve (solid line) with mean $\mu_G=0$ and standard deviation $\sigma_G=1$, with deviations only in the tails. Thus, $\epsilon_r$ closely follows a log-normal distribution over the accessible range of Rayleigh numbers. On the other hand, although the PDFs of $\chi_r$ are close to log-normal for $\Ray=10^7$, they are skewed for $\Ray=10^5$ because of the diffusion-dominated dynamics of temperature. These PDFs get closer to log-normal as $r$ is increased.}
\label{fig:PDFs}
\end{figure}
\begin{figure}
\captionsetup{width=1\textwidth}
  \centerline{\includegraphics[scale=0.19]{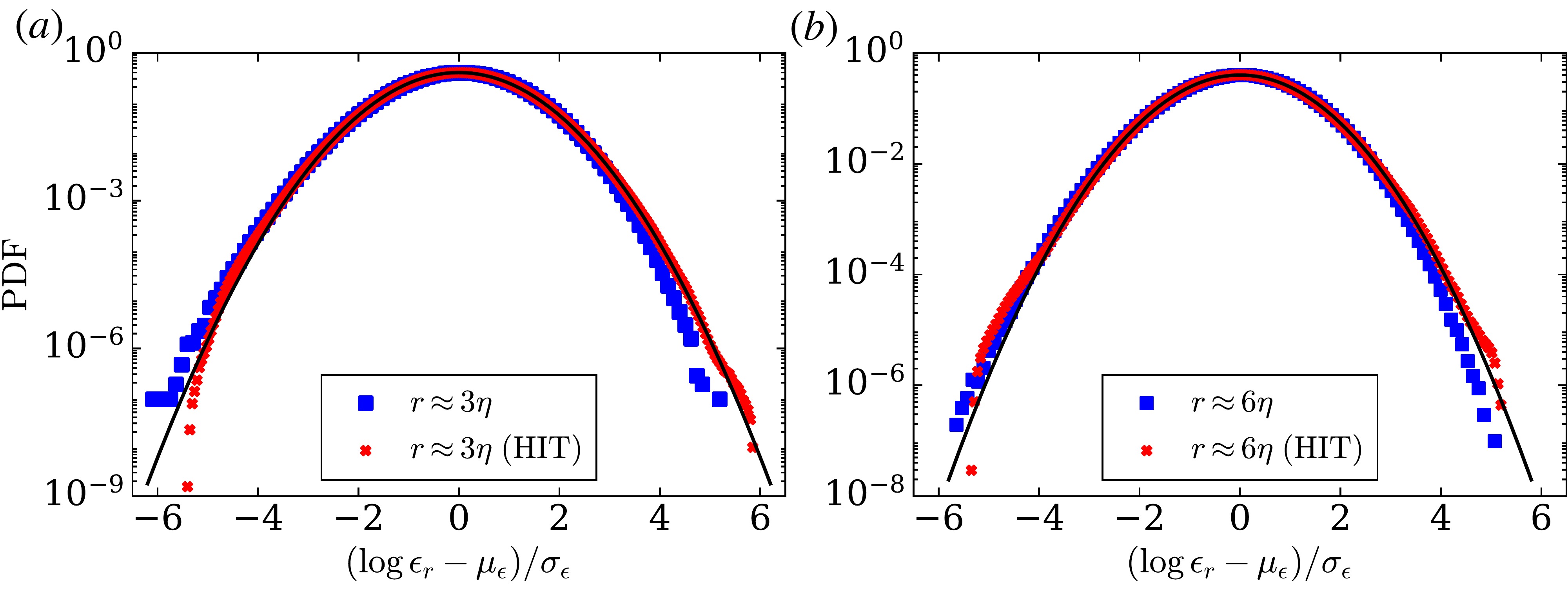}}
  \caption{Comparison of the PDFs of normalized viscous dissipation rates of thermal convection (blue) for $\Ray=10^7$ ($R_\lambda=709$) and homogeneous isotropic turbulence (red) for $R_\lambda=359$.  Panel ($a$) shows the PDFs of $\epsilon_r$ for $r=3 \eta$, and ($b$) shows the PDFs for $r = 6 \eta$. The PDFs of $\epsilon_r$ for thermal convection and homogeneous isotropic turbulence (HIT) are similar, closely following the log-normal distribution (black solid curves).}
\label{fig:PDFs_comparison_HIT}
\end{figure}
The dissipation rates were determined in the horizontal midplane in the bulk region ($5 \leq x/H \leq 20$ and $5 \leq y/H \leq 20$) away from the side walls which still gives a large area to draw samples of the fields for the analysis. We compute the probability density functions (PDFs) of $(\log \epsilon_r - \mu_\epsilon)/\sigma_\epsilon$ and $(\log \chi_r - \mu_\chi)/\sigma_\chi$ using the data from all the saved snapshots. In the above, $\mu_\epsilon$ and $\sigma_\epsilon$ are, respectively, the mean and standard deviation of $\log \epsilon_r$. Similarly, $\mu_\chi$ and $\sigma_\chi$ are the mean and standard deviation of $\log \chi_r$. The PDFs of the viscous dissipation rates are exhibited in figure~\ref{fig:PDFs}($a$) for $\Ray=10^5$ and in figure~\ref{fig:PDFs}($b$) for $\Ray=10^7$. In figures \ref{fig:PDFs}($c$) and \ref{fig:PDFs}($d$), we exhibit the PDFs of thermal dissipation rates for $\Ray=10^5$ and $\Ray=10^7$, respectively. It must be noted that for $\Ray=10^5$, we do not show the PDFs of $\chi_r$ with $r=4 \eta_C$ because, in this case, $r$ becomes larger than $H$ at the value of $4\eta_C$. For the viscous dissipation rate, the PDFs collapse and agree reasonably well with a Gaussian PDF (solid curve) with mean $\mu_G=0$ and standard deviation $\sigma_G=1$, implying that $\epsilon_r$ is log-normally distributed. On the other hand, while the PDFs of $\chi_r$ are close to log-normal for $\Ray=10^7$, they are highly skewed for $\Ray=10^5$, with the tails being fatter on the left side and sparser on the right. This skewness can be attributed to the fact that the temperature dynamics are diffusion-dominated for $\Ray=10^5$, due to which there is very poor mixing of $T$ in the bulk~\citep{Pandey:JFM2022}.
A careful examination of figure~\ref{fig:PDFs}(d) reveals that the PDFs of $\chi_r$ for $\Ray=10^7$ are also mildly skewed with extended left tails.
It is also observable in figure~\ref{fig:PDFs}(c) that the PDF for $r=2 \eta_C$ fits the log-normal curve better than that for $r=\eta_C$; implying that the distribution of $\chi_r$ becomes closer to log-normal as $r$ is increased. Thus, the usage of refined similarity hypothesis for determining the eddy diffusivity and hence the turbulent Prandtl number will be accurate only at large $r$.  

Log-normal PDFs of the coarse-grained viscous dissipation rate $\epsilon$ have been reported by \citet{Nakano2002} from DNS of homogeneous isotropic turbulence at a Taylor microscale Reynolds number of $R_{\lambda}=121$ with a nearly perfect collapse at the tails of the distribution, similar to the present data at $\Ray=10^7$. 
Therefore, we compare the PDFs of $\epsilon$ for $\Ray=10^7$ with those of a homogeneous isotropic turbulence simulation for the Taylor microscale Reynolds number $R_\lambda=359$. The details of the simulations are explained in Appendix~\ref{sec:HIT}.
We remark that the Taylor microscale Reynolds number in the horizontal midplane for $\Ray=10^7$ is $R_\lambda=709$, which is computed as
\begin{equation}
    R_\lambda = \left( \frac{25 \Ray}{9 \langle\epsilon\rangle^2 \Pran} \right)^{1/4} u_{rms}^2 \, ,
    \label{eq:Taylor_Re}
\end{equation}
where $u_{rms}$ is the root mean square velocity in the midplane, see \citet{Pandey:JFM2022}. 
We exhibit the PDFs of $\epsilon_r$ for thermal convection with $\Ray=10^7$ (shown in blue) and homogeneous isotropic turbulence (shown in red) in figure~\ref{fig:PDFs_comparison_HIT}(a) for $r=3 \eta$ and in figure~\ref{fig:PDFs_comparison_HIT}($b$) for $r = 6 \eta$.
The figures show that the PDFs of both thermal convection and homogeneous isotropic turbulence exhibit a similar behavior, closely following the log-normal distribution.

Log-normality of the scalar dissipation rate $\chi$ has been found in experiments by \cite{Sreenivasan:PF1977Temperatire} and in DNS by \cite{Schumacher:PF2005}. In the latter work, a fatter tail on the left hand side and a somewhat sparser one on the right hand side in comparison to a perfect log-normal distribution were detected, similar to our results on thermal dissipation rate. 
\section{Scale-resolved eddy viscosity and diffusivity} \label{sec:Results}

\subsection{Test of flux-gradient ansatz}
In this section, we discuss the behaviour of scale-dependent eddy viscosity, eddy diffusivity, and the turbulent Prandtl number.
\begin{figure}
\captionsetup{width=1\textwidth}
  \centerline{\includegraphics[scale=0.2]{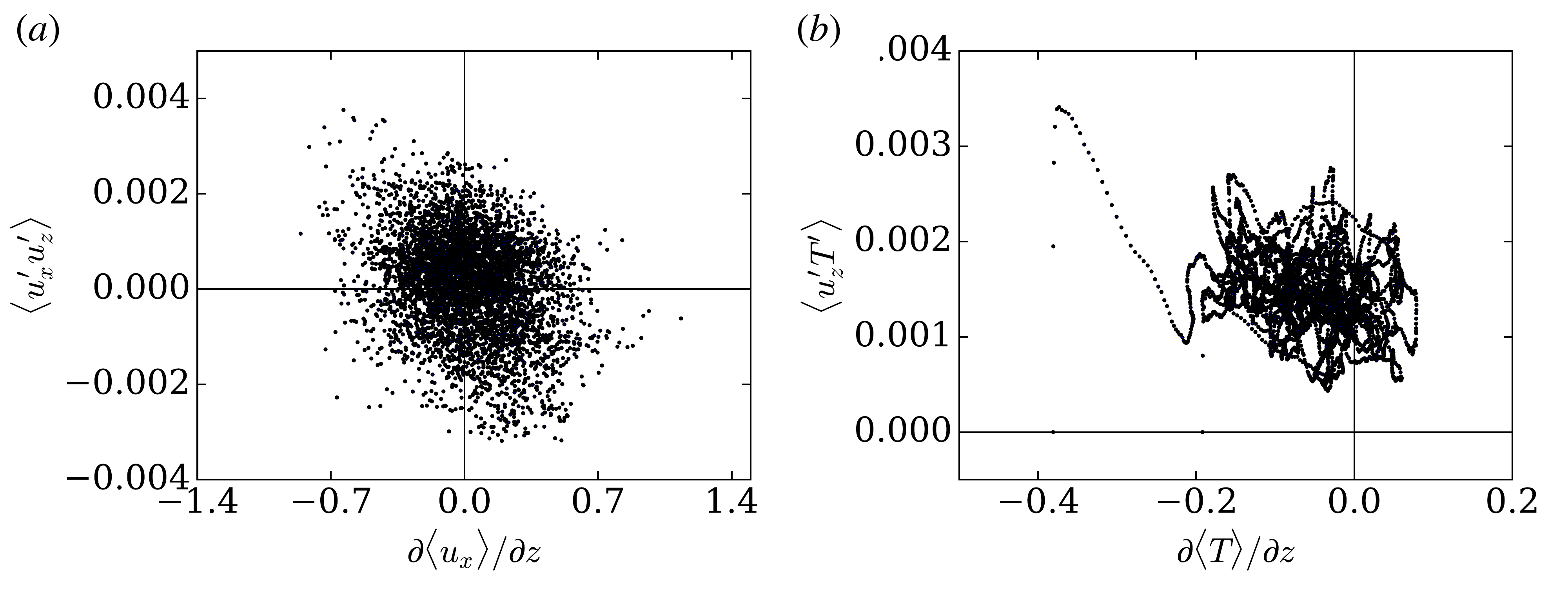}}
  \caption{Scatter plots for Rayleigh number $\Ray=10^7$: (a) Turbulent momentum flux $\langle u_x' u_z' \rangle_{x,t}$ as a function of the mean velocity gradient $\partial \langle u_x \rangle_{x,t}/\partial z$. (b) Turbulent convective heat flux $\langle u_z' T' \rangle_{x,t}$ as a function of the mean temperature gradient $\partial \langle T \rangle_{x,t}/\partial z$. The velocity field exhibits strong fluctuations, resulting in a nearly homogeneous distribution of the phase points and thus making it challenging to estimate the eddy viscosity using the flux gradient method. Data are obtained in the midplane by a combined average with respect to $x$ direction and time $t$. A total of 5120 data points taken from 10 data snapshots are plotted.}
\label{fig:Velocity_profile}
\end{figure}
The eddy viscosity and diffusivity are frequently estimated using the flux-gradient relations, see, e.g., \citet{Wilcox:book:CFD}. Accordingly, the eddy viscosity $\nu_e$ is estimated by dividing the Reynolds stress $-\langle u_i' u_j' \rangle$ by the time-averaged velocity gradient $\partial \langle u_i\rangle/\partial x_j$, and the eddy diffusivity $\kappa_e$ is estimated by dividing the turbulent heat flux $-\langle u_j'T' \rangle$ by the time-averaged temperature gradient $\partial \langle T\rangle/\partial x_j$, as discussed in \S~\ref{sec:Introduction} (see equations~\eqref{eq:bouss}).

However, in RBC, there is an absence of a mean flow~\citep{Samuel:JFM2024}; mean velocity and temperature gradients in the bulk are practically zero or indefinite. This aspect is exhibited in figure~\ref{fig:Velocity_profile}($a$), where we plot the turbulent momentum flux $\langle u_x' u_z' \rangle_{x,t}$ versus the mean velocity gradient $\langle \partial \langle u_x \rangle_{x,t}/\partial z$, and in figure~\ref{fig:Velocity_profile}($b$), where we plot the turbulent heat flux $\langle u_z' T' \rangle_{x,t}$ versus the mean temperature gradient $\langle \partial \langle T \rangle_{x,t}/\partial z$. Here, $\langle \cdot \rangle_{x,t}$ represents averaging over time $t$ and the $x$-direction, and the quantities are computed in the horizontal midplane. The phase points in figure~\ref{fig:Velocity_profile}($a$) are distributed homogeneously in all the four quadrants, implying that the turbulent viscosity $\nu_e = -\langle u_x' u_z' \rangle_{x,t}/(\partial \langle u_x \rangle_{x,t}/\partial z)$ frequently changes sign. In figure~\ref{fig:Velocity_profile}($b$), the data points are distributed mostly in the second quadrant, implying that the turbulent diffusivity $\kappa_e = -\langle u_z' T' \rangle_{x,t}/(\partial \langle T \rangle_{x,t}/\partial z)$ is largely positive. This is expected since $u_z'$ and $T'$ are positively correlated~\citep{Verma:NJP2017}, and the temperature decreases weakly with $z$ in the bulk region for small Prandtl numbers~\citep{Pandey:PRF2021,Pandey:PD2022,Pandey:JFM2022}. However, some of the data points lie in the first quadrant, resulting in several instances of positive temperature gradient and hence negative turbulent diffusivity. Thus, computing the eddy quantities using the flux-gradient method becomes highly challenging for RBC, and therefore , we resort to the alternative approach of refined similarity hypothesis for computing these quantities.

\subsection{Application of self-similarity framework}
Having verified the approximate log-normal distributions of $\epsilon_r$ and $\chi_r$, we proceed to compute the eddy viscosity and eddy diffusivity using equations~(\ref{eq:ScaleEViscosity}) and (\ref{eq:ScaleEDiff}), respectively, for scales $r\in [\eta, H/2]$ for $\Ray=10^6$ and $10^7$, and for $r \in [\eta,H]$ for $\Ray=10^5$. Once again, the aforementioned quantities are computed using the data in the bulk region ($5 \leq x/H \leq 20$ and $5 \leq y/H \leq 20$) on the horizontal midplane and are averaged over all time frames. 
It must be noted that we get similar results on computing the eddy quantities in other planes close to the horizontal midplane (see Appendix~\ref{sec:OtherPlanes}). 
Here, we remark that for $r>H/2$, the expressions for the eddy viscosity and diffusivity given by (\ref{eq:ScaleEViscosity}) and (\ref{eq:ScaleEDiff}) lose accuracy, due to the effects of the walls; thus, we restrict distances to $r\le H/2$. However, for $\Ray=10^5$, the Corrsin length scale itself is $\eta_C\approx 0.5$ (as shown in table~\ref{table:Simulation}). Thus, we had to go up to distance of $r\sim H$ in order to capture an adequate range of length scales for which $\kappa_e>\kappa$.
\begin{figure}
\captionsetup{width=1\textwidth}
  \centerline{\includegraphics[scale=0.38]{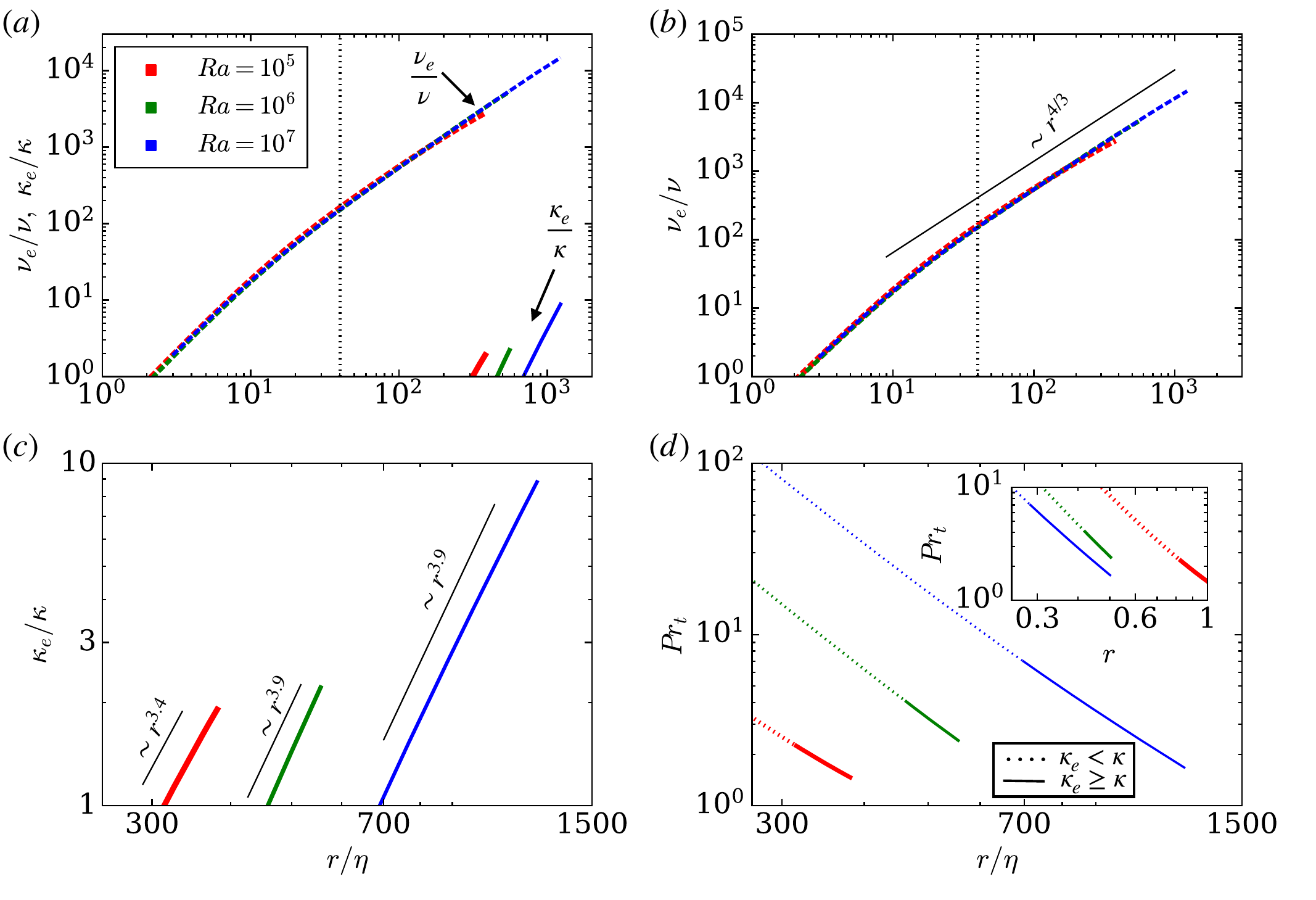}}
  \caption{Plots of eddy viscosity, eddy diffusivity, and scale-resolved turbulent Prandtl number versus the normalized length scale $r/\eta$ for $\Ray=10^5$ (red lines), $\Ray=10^6$ (green lines), and $\Ray=10^7$ (blue lines) in decreasing order of thickness. (a) Normalized eddy viscosity $\nu_e/\nu$ and eddy diffusivity $\kappa_e/\kappa$. (b) Magnified plot of $\nu_e/\nu$ which fits closely with $r^{4/3}$ curve. (c) Magnified plot of $\kappa_e/\kappa$ along with the corresponding best-fit curves. (d) Scale-dependent turbulent Prandtl number $\Pran_t$. In panels (a) and (b), the vertical dotted line represents $r/\eta=40$, which marks the lower end of the inertial subrange. In panel (d), the dotted lines correspond to the regime where $\kappa_e<\kappa$, and the solid lines correspond to $\kappa_e \geq \kappa$. The inset in (d) exhibits the plots of $\Pran_t$ versus the length scale $r$ for $0.25 \leq r \leq 1$.}
\label{fig:Eddy}
\end{figure}

We normalize the eddy viscosity and diffusivity with their respective molecular counterparts and plot these quantities versus $r/\eta$ in figure~\ref{fig:Eddy}($a$). It must be noted that magnitudes of the eddy viscosity and diffusivity which are smaller than their molecular counterparts do not make sense as $\kappa_e < \kappa$ effectively means the presence of temperature filaments which are smaller than the Corrsin scale. Therefore, we exhibit only those regimes where $\kappa_e \geq \kappa$. Since we consider $\Pran =0.001 \ll 1$ and moderate Rayleigh numbers, the range of length scales for which $\kappa_e \geq \kappa$ is small. Hence, only a very small portion of $\kappa_e$ curves is exhibited in figure~\ref{fig:Eddy}($a$) and there is no overlapping regime of $r/\eta$ corresponding to $\kappa_e \geq \kappa$ for the three Rayleigh numbers. 

Figure~\ref{fig:Eddy}(a) shows several interesting features. Both eddy viscosity and diffusivity increase with the length scale, which is expected. The curves of the normalized eddy viscosity collapse very well for all the Rayleigh numbers, implying that the normalized eddy viscosity depends only on the normalized length scale, and not on $\Ray$. However, the normalized eddy diffusivity curves do not collapse, suggesting that these quantities do not depend on the normalized length scale alone, and there is an additional $\Ray$-dependence. We have verified that a rescaling with $\eta_C$ rather than $\eta$ did not lead to a collapse of the data. The trend of the curves for $\kappa_e$ implies that for a particular $r/\eta$, the normalized eddy diffusivity decreases with the increase of $Ra$. However, since there is no overlapping regime of $r/\eta$ corresponding to $\kappa_e>\kappa$, the $\Ray$-dependence of the normalized eddy diffusivity remains inconclusive. 

\begin{figure}
\captionsetup{width=1\textwidth}
  \centerline{\includegraphics[width=0.6\textwidth]{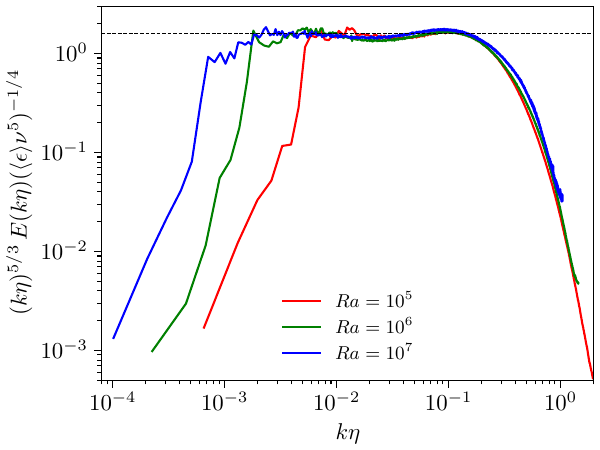}}
  \caption{Normalized kinetic energy spectra versus the normalized wavenumber $k\eta$ in the midplane exhibit an inertial subrange, which broadens with increasing $\Ray$. The end of the plateau region at $k\eta \approx 0.15$ corresponds to the lower limit of the inertial subrange at $r/\eta \approx 40$. The horizontal line indicates the Kolmogorov constant $K_K \approx 1.6$.}
\label{fig:Eu_k}
\end{figure}

Figure~\ref{fig:Eddy}(b) reveals that the eddy viscosity curves fit well with $\nu_e \sim r^{4/3}$ for $r \gg \eta$, which is consistent with the effective eddy viscosity relation given by equation~\eqref{eq:EffectiveScaleEViscosity}. As mentioned in \S~\ref{sec:RefinedSimilarityVelocity}, this scaling relation is expected to hold in the inertial subrange, which can be determined by investigating the kinetic energy spectrum $E(k)$~\citep{Verma:NJP2017}. \citet{Pandey:JFM2022} showed that $E(k)$ in low-$Pr$ mesoscale convection follows the Kolmogorov scaling, i.e., $E(k) \sim k^{-5/3}$~\citep{Kolmogorov:DANS1941Structure}, where $k$ is the wavenumber. The spectrum is obtained in the midplane with respect to all three velocity components, i.e., $k=(k_x^2+k_y^2)^{1/2}$. In figure~\ref{fig:Eu_k}, we display the normalized kinetic energy spectra in the midplane by plotting $ (k\eta)^{5/3} \, E(k\eta) (\langle\epsilon\rangle \nu^5)^{-1/4}$ against the normalized wavenumber $k\eta$. A plateau---which becomes wider with increasing $\Ray$---can be detected for all the Rayleigh numbers. The right hand side end of the plateau is observed at $k\eta \approx 0.15$ for all the Rayleigh numbers, which provides an estimate for the lower end of the inertial subrange. This corresponds to $r/\eta \approx 40$ in figures~\ref{fig:Eddy}(a) and (b), which is indicated by a vertical dotted line. It is clear that the scaling $\nu_e \sim r^{4/3}$ in figure~\ref{fig:Eddy}(b) is observed in the inertial subrange, which extends up to $r/\eta = 850, 1550$, and $3150$ for $Ra = 10^5, 10^6$, and $10^7$, respectively.

As exhibited in figure~\ref{fig:Eddy}($c$), the best fit curves for eddy diffusivity follow $\kappa_e \sim r^{3.9}$ for $\Ray=10^6$ and $10^7$, and $\kappa_e \sim r^{3.4}$ for $\Ray=10^5$. Comparing these fits with the derived expression for effective thermal diffusivity given by equation~(\ref{eq:ScaleEDiff}), we infer that the prefactor $\langle w_x^2 w_T^2 \epsilon_r^{1/3} \rangle$ has a strong dependence on the length scale $r$. This shows that the scaling which we developed in \S~\ref{sec:ScalarFields} is not detected for the present data. We also remark that the observed difference in the exponents between $\Ray=10^5$ and $\Ray>10^5$ might be due to the fact that the length scales analyzed for $\Ray=10^5$ are well above $0.5H$ and the effects of boundaries become relevant. It remains to be seen how the results change when higher Rayleigh numbers at this low $Pr$ are accessible.  

\begin{figure}
\captionsetup{width=1\textwidth}
  \centerline{\includegraphics[width=1\textwidth]{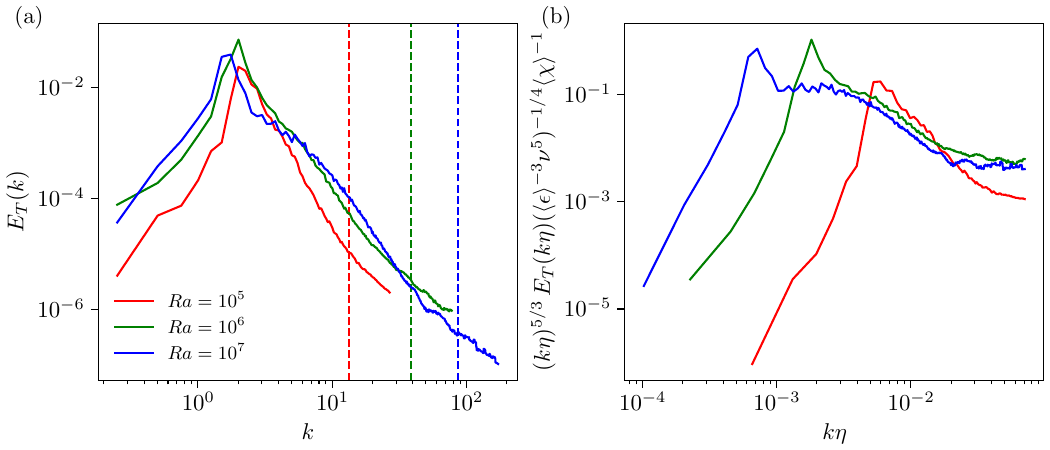}}
  \caption{(a) Thermal energy spectra in the midplane decay rapidly beyond the prominent maximum that corresponds to the characteristic scale of superstructures. Dashed vertical lines indicate the Corrsin wavenumber $2 \pi/\eta_C$. (b) Normalized spectra do not exhibit an inertial-convective subrange, except for $\Ray = 10^7$, where a narrow plateau region is detected for $0.001 \leq k\eta \leq 0.003$. }
\label{fig:ET_k}
\end{figure}

To gain further insights, we look at the thermal energy spectra $E_T(k)$ in the midplane, which is defined as 
\begin{equation}
    E_T(k) = \pi k |\hat{T}(k)|^2 \, ,
\end{equation}
where $\hat{T}(k)$ is the Fourier transform of the temperature field in the midplane;
see \citet{Pandey:JFM2022} for further details. The spectra $E_T(k)$ are displayed in figure~\ref{fig:ET_k}(a), where the maximum wavenumbers are restricted up to twice the value corresponding to the Corrsin scale. The Corrsin wavenumber $2\pi/\eta_C$ is indicated by dashed vertical lines. There is a prominent maximum in $E_T(k)$ around $k \approx 2$, beyond which a rapid decay of the spectrum is observed. The maxima correspond to the characteristic scales of the turbulent superstructures~\citep{Pandey:NC2018}, and the rapid decline thereafter is consistent with a highly diffusive temperature field observed in very-low-$Pr$ convection, see \citet{Pandey:JFM2022} for visualizations of the temperature field for $Pr = 10^{-3}$. 

It has been proposed that in the inertial-convective subrange, the thermal energy spectrum scales as~\citep{Corrsin:JAP1951, Sreenivasan:PF1996}
\begin{equation}
    E_T(k) = K_C \langle\epsilon\rangle^{-1/3} \langle\chi\rangle k^{-5/3} \, ,
\end{equation}
which in the normalized form reads
\begin{equation}
    E_T(k\eta) = K_C (\langle\epsilon\rangle^{-3} \nu^5)^{1/4} \langle\chi\rangle (k\eta)^{-5/3}.
\end{equation}
Here $K_C$ is the Corrsin-Obukhov constant~\citep{Sreenivasan:PF1996}. In figure~\ref{fig:ET_k}(b), we show the normalized thermal energy spectrum and observe that the inertial-convective subrange, where the spectrum exhibits a plateau, hardly exists. Only for $\Ray = 10^7$, a narrow plateau region could be detected for $k\eta \in [0.001 - 0.003]$, the upper end of which corresponds roughly to $r/\eta \approx 2000$. This scale is beyond the maximum $r/\eta$ exhibited in figure~\ref{fig:Eddy}(c,d).

Since the eddy diffusivity increases more steeply with $r$ compared to the eddy viscosity, the scale-dependent turbulent Prandtl number decreases with increasing $r$. This is exhibited in figure~\ref{fig:Eddy}($d$), where we plot $\Pran_t$ versus the normalized length scale $r/\eta$, focusing on the regions corresponding to $\kappa_e > \kappa$. To this end, the curve segments of $\Pran_t(r)$ are plotted as solid lines for $\kappa_e \geq \kappa$. Although $\Pran_t$ physically does not exist for $\kappa_e < \kappa$, the plots of $\Pran_t$ are still shown in this regime as dotted lines. This is done primarily to understand the trend of $\Pran_t$ had the regime corresponding to $\kappa_e \geq \kappa$ been wider. Figure~\ref{fig:Eddy}($d$) shows that $\Pran_t$ takes values between 1 and 10, which is consistent with earlier studies on low-$\Pran$ flows~\citep{Reynolds:IJHMT1975,Jischa:IJHMT1979,Bricteux:NED2012,Pandey:PRF2021,Pandey:PD2022}. This is unlike the case for moderate- and large-$\Pran$ fluids where $\Pran_t$
is less than unity. Thus, the scale-resolved turbulent Prandtl number is by almost four orders of magnitude higher than its molecular counterpart for the present case at very low $\Pran$.

Like the eddy diffusivity, the turbulent Prandtl number shows additional dependence on $\Ray$ apart from $r/\eta$. Again, a conclusive $\Ray$-dependence of $\Pran_t$ cannot be obtained because of the lack of an overlapping regime corresponding to $\kappa_e \geq \kappa$. However, going by the trend of the turbulent Prandtl number as shown by dotted lines, it can be suggested that for a particular $r/\eta$, $\Pran_t$ increases with increasing $\Ray$. This is consistent with the fact that the normalized eddy diffusivity decreases with an increase in  $\Ray$ as explained earlier.
However, on a particular length scale $r$, $\Pran_t$ decreases with the increase of $\Ray$; this is exhibited in the inset of figure~\ref{fig:Eddy}(d) where we plot $\Pran_t$ versus $r$. This behaviour of $\Pran_t$ is conclusive for $\Ray=10^6$ and $10^7$, where there is a small overlapping regime of $r$ corresponding to $\kappa_e \geq \kappa$. This trend is also consistent with results of \citet{Pandey:PRF2021, Pandey:PD2022}, who computed $\Pran_t$ using the $k_u$-$\epsilon$ approach. In a future work, we plan to extend our studies to a wider range of Rayleigh and Prandtl numbers to examine the universality of the observed trends of $\Pran_t$ in the present work.

\section{Conclusions and outlook}
In this paper, we analyzed the scale-dependent turbulent Prandtl number $\Pran_t$ of low-$\Pran$ mesoscale convection using the refined similarity framework of \citet{Kolmogorov:JFM1962} and \citet{Obukhov:JFM1962}. This hypothesis implies that the coarse-grained viscous and thermal dissipation rates are log-normally distributed in the bulk, which we verified using our data obtained from DNS of RBC for $\Pran=10^{-3}$ and $\Ray=10^5$, $10^6$, and $10^7$. We computed the scale-dependent eddy viscosity $\nu_e$ and eddy diffusivity $\kappa_e$ using the refined similarity hypothesis, and obtained $\Pran_t$ as $\nu_e/\kappa_e$. The log-normal distribution of the kinetic energy dissipation rate in the bulk of the convection flow was found to agree well with that of three-dimensional homogeneous isotropic turbulence in a similar range of Taylor microscale Reynolds numbers.

We showed that both $\nu_e$ and $\kappa_e$ increase with the increasing length scale $r$. The curves of the eddy viscosity normalized by the molecular viscosity collapse well when plotted against $r/\eta$, implying that the normalized eddy viscosity is a function of $r/\eta$ only. We further found that $\nu_e$ scales as  $r^{4/3}$, which is consistent with \citet{Kolmogorov:JFM1962}, \citet{Obukhov:JFM1962} and \citet{Yakhot:PD2006}. On the other hand, the curves of the eddy diffusivity normalized by the molecular thermal diffusivity do not collapse when plotted against $r/\eta$, and exhibit a steeper scaling ranging from $\kappa_e \sim r^{3.4}$ for $\Ray=10^5$ to $\kappa_e \sim r^{3.9}$ for $\Ray=10^7$. Thus, $\kappa_e/\kappa$ has an additional $\Ray$ dependence. Since the slopes of the normalized eddy diffusivity are steeper compared to those of the normalized eddy viscosity, $\Pran_t$ decreases with an increase of $r$ for all the explored $\Ray$. 

As $\Pran$ in the present work is very low, thermal diffusion contributes significantly in the heat transfer through the fluid, resulting in large Corrsin scales of the order the domain size. Hence, the range of scales for which $\kappa_e \geq \kappa$ is narrow, and we do not get overlapping regimes of $r/\eta$ for which the eddy diffusivity is greater than its molecular counterpart. However, based on the trend of the curves for $\kappa_e$ and $\Pran_t$, it can be surmised that for a particular $r/\eta$, $\kappa_e$ increases, while $\Pran_t$ decreases, with a decrease of $\Ray$. Moreover, at a particular length scale $r$, the $\Pran_t$ is expected to increase with decreasing $\Ray$. The turbulent Prandtl number takes values between 1 and 10 at all scales pertaining to $\kappa_e\geq \kappa$, which is unlike the case for moderate- and large-$\Pran$ fluids, where $\Pran_t$ is slightly less than unity. This is consistent with the fact that for small $\Pran$, the turbulent transport of momentum exceeds that of heat, resulting in larger turbulent Prandtl numbers~\citep{Pandey:PRF2021}. Our approach demonstrated that vigorous turbulence in the bulk of a low-Prandtl-number convection layer behaves effectively as a high-Prandtl-number flow. This confirms previous discussions by \citet{Omara2016} and in particular by \citet{Bekki:ApJ2017} in the context of solar convection. This is shown here for turbulent convection in its simplest form, as a Boussinesq setup. 

The results of this work provide valuable insights into the behaviour of eddy dissipation and the resulting scale-dependent turbulent Prandtl number for very-low-$\Pran$ thermal convection.  Although we explored a small set of parameters, the trends of the quantities analysed in this work are expected to be valid over a wide range of $\Ray$ and $\Pran$. Our results are expected to provide important inputs for developing accurate subgrid models for low-$\Pran$ convection, which, in turn, will be helpful for modeling flows in nature and technology.


\backsection[Acknowledgements]{The authors gratefully acknowledge the Gauss Centre for Supercomputing e.V. (https://www.gauss-centre.eu) for funding this project by providing computing time on the GCS Supercomputer SuperMUC-NG at Leibniz Supercomputing Centre (https://www.lrz.de). S.B. thanks Param Himalaya of Indian Institute of Technology Mandi for providing resources to carry out postprocessing computations. A.P.
acknowledges financial support from SERB, India under the grant SRG/2023/001746. The work of T.G. was supported by JSPS KAKENHI Grant No.25K01161 and used computational resources of Fugaku provided by RIKEN  through the HPCI System Research Project (Project ID: hp250029). We thank Katepalli Sreenivasan for helpful discussions.}


\backsection[Declaration of interests]{ The authors report no conflict of interest.}

\backsection[Data availability statement]{The data that support the findings of this study are available on reasonable request.}

\backsection[Author ORCIDs]{\\
S. Bhattacharya, https://orcid.org/0000-0001-7462-7680;\\
D. Krasnov, https://orcid.org/0000-0002-8339-7749;\\
A. Pandey, https://orcid.org/0000-0001-8232-6626;\\
T. Gotoh, https://orcid.org/0000-0002-9668-582X;\\
J. Schumacher, https://orcid.org/0000-0002-1359-4536.}

\backsection[Author contributions]{All authors designed the research. S.B., D.K., A.P. and T.G. conducted the numerical analysis. All authors discussed the results and wrote the manuscript.}

\appendix
\section{Details of the simulations of homogeneous isotropic turbulence} \label{sec:HIT}

The velocity of an incompressible fluid of unit density obeys the Navier-Stokes equation 
and is excited by the solenoidal Gaussian random force ${\mbox{\boldmath $f$}}$ 
which has the spectral support 
at the low wavenumber range~\citep{Gotoh:PRL2023} and  
follows the Ornstein-Uhlenbeck process with the characteristic time of $O(1)$. 
Direct numerical simulations are performed by  
using the pseudo-spectral method in space with the number of grid points 
$N=2048^3$ and the 4th-order Runge-Kutta-Gill method in time. 
This simulation is a continuation of RUN 2048C described in the supplemental material 
of \citet{Gotoh:PRL2023}. 
In the statistically steady state, the time average of the PDF is taken 
over the duration of $11.7T_e$, where $T_e$ is the large eddy turn over time, 
and the samples are gathered every $0.038T_e$ time unit.    
The Taylor microscale Reynolds number is $R_\lambda=359$ and $\nu=0.0003,\ \langle\epsilon\rangle=0.331$ 
and $k_{max}\eta=2.93$.

\section{Eddy quantities and turbulent Prandtl number on other planes close to the horizontal midplane} \label{sec:OtherPlanes}
Thus far, we have computed the scale-resolved eddy viscosity, eddy diffusivity, and the turbulent Prandtl number using the numerical data in the horizontal midplane.
In this section, we verify the consistency of our results in the bulk region by computing and comparing the aforementioned quantities in other horizontal planes close to the midplane.

\begin{figure}
\captionsetup{width=1\textwidth}
  \centerline{\includegraphics[scale=0.38]{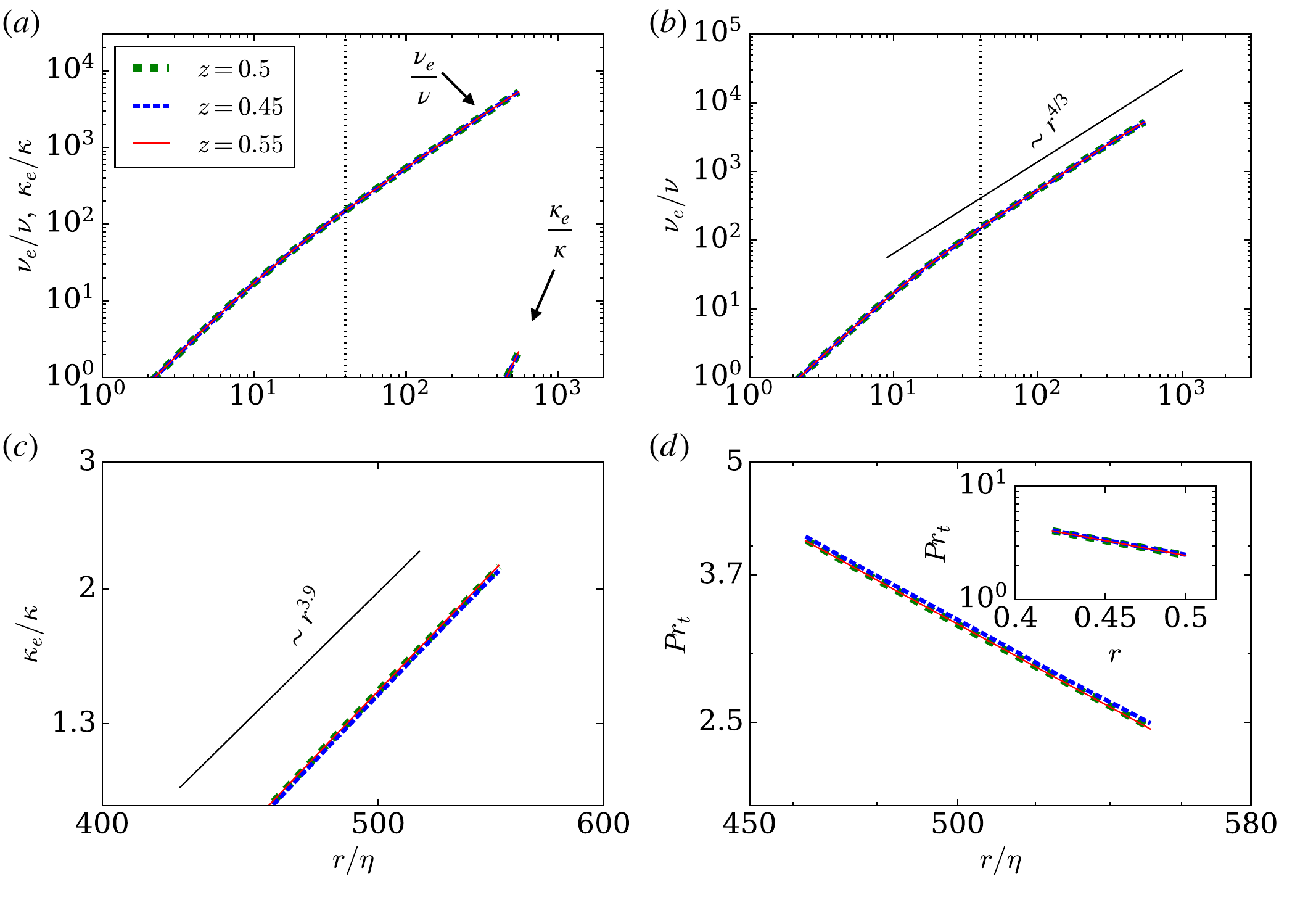}}
  \caption{For $\Ray=10^6$: Plots of eddy viscosity, eddy diffusivity, and turbulent Prandtl number versus the normalized length scale $r/\eta$ computed on $z=0.5$ (dashed green lines), $z=0.45$ (dashed blue lines), and $z=0.55$ (solid red lines). (a) Normalized eddy viscosity $\nu_e/\nu$ and eddy diffusivity $\kappa_e/\kappa$. (b) Magnified plot of $\nu_e/\nu$ which fits closely with $r^{4/3}$ curve in the inertial subrange. (c) Magnified plot of $\kappa_e/\kappa$ along with the corresponding best-fit curve. (d) Scale-dependent turbulent Prandtl number $\Pran_t$. The vertical dotted line in panels (a) and (b) represents $r/\eta=40$, which marks the lower end of the inertial subrange. The inset in (d) exhibits the plots of $\Pran_t$ versus the length scale $r$. The turbulent quantities computed on the three planes match closely with each other.}
\label{fig:Eddy_DiffPlanes}
\end{figure}

We compute the scale-resolved eddy viscosity and eddy diffusivity using equations~(\ref{eq:ScaleEViscosity}) and (\ref{eq:ScaleEDiff}), respectively, for scales $r\in [\eta, H/2]$ and $\Ray=10^6$, using the data on planes $z=0.45H$ and $z=0.55H$. 
The plots of the normalised eddy viscosity ($\nu_e/\nu$) and eddy diffusivity ($\kappa_e/\kappa$), computed on the aforementioned planes, versus the normalised length scale $r/\eta$ are exhibited in figure~\ref{fig:Eddy_DiffPlanes}($a$).
The eddy quantities computed using the data on the horizontal midplane are also plotted in the figure for reference.
The figure shows that the eddy quantities computed on these different planes are similar, implying that our results are consistent in the bulk region. Figure~\ref{fig:Eddy_DiffPlanes}($b$) shows that for $r \gg \eta$, the curves of eddy viscosity computed on all the planes fit well with $\nu_e(r) \sim r^{4/3}$, whereas as per figure~\ref{fig:Eddy_DiffPlanes}($c$), the curves of eddy diffusivity fit well with $\kappa_e(r) \sim r^{3.9}$.
The turbulent Prandtl number computed on these planes are also very close to each other as exhibited in figure~\ref{fig:Eddy_DiffPlanes}($d$).

We further quantify the similarities in the turbulent quantities computed on the three horizontal planes.
We compute the percentage deviation $\Delta$ between the quantities $\phi=\{\nu_e, \kappa_e, \Pran_t\}$ computed in the three planes as follows
\begin{equation}
    \Delta_{0.45} = 100 \frac{|\phi_{0.45} - \phi_{0.5}|}{\phi_{0.5}}, \quad
    \Delta_{0.55} = 100 \frac{|\phi_{0.55} - \phi_{0.5}|}{\phi_{0.5}},
    \label{eq:Deviation}
\end{equation}
where the subscripts 0.45 and 0.55 correspond to $z=0.45H$ and $z=0.55H$, respectively.
The deviations are within 0.6\% for $\nu_e$, 1.3\% for $\kappa_e$, and 1.4\% for $\Pran_t$.
Thus, the results computed on all the planes are similar, hence implying that our results are consistent in the bulk.

\bibliographystyle{jfm}

\end{document}